# AdaPhish: AI-Powered Adaptive Defense and Education Resource Against Deceptive Emails

Rei Meguro
*College of Engineering*
*University of Michigan-Ann Arbor*
Michigan, United States
rmeguro@umich.edu

Ng S. T. Chong
*Campus Computing Centre*
*United Nations University*
Tokyo, Japan
ngstc@unu.edu

*Abstract*— **Phishing attacks remain a significant threat in the digital age, yet organizations lack effective methods to tackle phishing attacks without leaking sensitive information. Phish bowl initiatives are a vital part of cybersecurity efforts against these attacks. However, traditional phish bowls require manual anonymization and are often limited to internal use. To overcome these limitations, we introduce AdaPhish, an AI-powered phish bowl platform that automatically anonymizes and analyzes phishing emails using large language models (LLMs) and vector databases. AdaPhish achieves real-time detection and adaptation to new phishing tactics while enabling long-term tracking of phishing trends. Through automated reporting, adaptive analysis, and real-time alerts, AdaPhish presents a scalable, collaborative solution for phishing detection and cybersecurity education.**

*Keywords—phish bowl, phishing detection, large language model, anonymization*

## I. INTRODUCTION

Phishing email scams, also known as phish, are a type of cybersecurity attack where attacker deceives users into either revealing sensitive information such as passwords, credit card numbers, or personal data, or opening malicious attachments by disguising as a trustworthy entity. The 2024 Verizon Data Breach Investigation Report indicates that 68% of breaches involved human interaction, including phishing attacks [1]. With AI-driven scams becoming harder to detect, filtering and documenting phishing attempts is increasingly vital.

Phish bowls are a collaborative platform where users can submit known phishes, helping cybersecurity teams and researchers identify and analyze common phishing tactics. In this paper, we extend the idea of the phish bowl by introducing AdaPhish, a collaborative platform powered by large language models (LLM) and a vector database. AdaPhish not only enables the submission and documentation of past phishing emails but also learns from the data to adapt to emerging phishing tactics and provide real-time alerts for potential phishing campaigns. In addition to serving as an adaptive phishing detection mechanism, AdaPhish offers a comprehensive repository of past phishing emails, which can be leveraged by security teams and organizations for educational purposes, trend analysis, and the creation of phishing simulations.

## II. CONTRIBUTIONS

Our proposed platform, AdaPhish, offers several distinct advantages over traditional phishing detection systems:

### A. Reporting Mechanism

Unlike traditional phish bowls, which require manual anonymization of sensitive information, AdaPhish uses GPT-4o to automatically replace named entities while retaining key context. This ensures confidential information is protected without user intervention, enabling faster and more efficient email submissions, and making the platform beneficial for both internal and external teams.

### B. Automated Analysis

AdaPhish employs two AI models to enhance detection accuracy. One model uses embeddings to compare emails and detect new phishing tactics, while the other assesses risks based on pre-trained knowledge. This dual approach allows the platform to instantly adapt to new threats, even in the early stages when fewer samples are available.

### C. Real-Time Updates

AdaPhish continuously tracks phishing trends, providing real-time alerts for emerging campaigns. This allows organizations to act swiftly, preventing large-scale attacks and giving users timely instructions on handling phishing attempts.

### D. Enhanced Search and Retireval

By using vector embeddings, AdaPhish enables robust search functionality, allowing natural language queries and semantic grouping of emails. This enhances its ability to handle variations like synonyms or visually similar characters, improving the usability of the phish bowl.

## III. PLATFORM ARCHITECTURE

AdaPhish is composed of three core components: the phish bowl database, a front-end web app, and a backend API. Emails submitted are processed by the backend and stored in the database. The front-end serves as an interface to access the system's security and educational tools. Fig. 1 illustrates the platform's architecture. Key defensive and educational features of the platform include:

**Detection Interface:** Accepts email text or images, analyzes them using a pretrained LLM, and classifies them as "phishing" or "benign" with confidence levels.

**Email Sharing:** Allows users to submit known phishing emails, which are added to the database to improve detection.

**Realtime Alerts:** Detects repeated phishing patterns and sends alerts to warn users of potential widespread campaigns.

**Natural Language Search:** Enables searching for anonymized phishing emails using natural language, improving the discovery of related emails.

**Trend Reporting:** Provides insights into the most frequently submitted phishing emails, helping users stay updated on emerging scams.

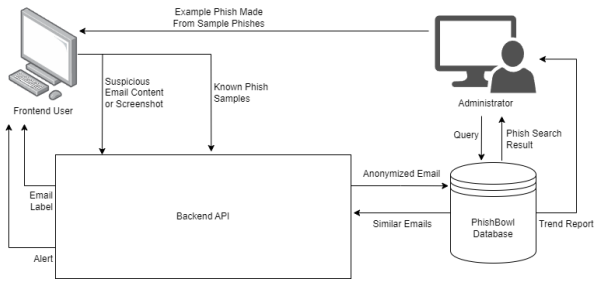

Fig. 1. Platform Architecture

IV. DETECTION MECHANISM

This section describes each component of the finalized detection mechanism as illustrated in Fig. 2. The detection performances of previous iterations building up to the finalized architecture is explored later in section **V.B**.

*A. OCR Text Extractor*

The text extractor is the first step of the detection mechanism, responsible for extracting the email content, comprised of the body, subject, and sender, from an email screenshot. This is done by using Google's Tesseract OCR engine [2] to extract and generate a Pandas data frame, containing text locations, contents, confidence levels (0 to 100), and heights. The following steps are then applied:

**Confidence Filtering:** The data frame is filtered to only keep words with a confidence score above a threshold, $t_{OCR}$. Words that don't meet this threshold are discarded, ensuring higher accuracy. The remaining text is grouped by lines, and their average text height is calculated.

**Header Detection:** Words commonly found in the email header (e.g., "from," "to," "subject," "sender") are matched using a regular expression. The header is defined as all lines up to the last occurrence of these header terms or a cutoff $t_{header}$ to avoid capturing the entire email. The variable `header_until` is calculated as the smaller of these two values, marking the end of the header.

**Body Detection:** Similar to the header, the email body is identified by searching for common greeting words (e.g., "hi," "hello") after the header. The variable `body_from` is set to the first line where a match occurs, or to 0 if no match is found. This potential overlap between the header and body ensures no important lines are excluded. The body is then set as all lines starting from `body_from`.

**Subject Extraction:** The subject is extracted either via regular expressions (looking for "subject:") or by selecting lines where the average text height is greater than the median line height by a factor of $k_{subject}$. However, texts with height greater than $t_{logo}$ is ignored to prevent misidentifying logos or images as the subject.

**Sender Extraction:** The sender is identified by matching text that looks like an email address through another regular expression.

The process is designed to handle typical email formats seen in services such as Gmail and Outlook for both mobile and desktop versions. The values $t_{OCR} = 80, t_{header} = 7, k_{subject} = 1.25$ and $k_{logo} = 1.5$, respectively, were shown to work the best during testing.

*B. Email Content Anonymizer*

The anonymizer masks sensitive information, like names and companies, while preserving essential details to help identify impersonation. For example, terms like "HR" or "Microsoft" remain unaltered to support detection. To achieve this, chain-of-thought prompting, a strategy that improves reasoning for complex tasks by laying out detailed instructions and examples [3], was applied to GPT-4o, guiding the model on which entities to anonymize and which to retain. Template. I was used as the prompt to accomplish anonymization:

| Template. I. Email Anonymization |
|---|
| I want you to act as an email anonymization toolkit to help mask sensitive information from emails submitted by the user. The input will be text content, sectioned by subject, sender, and body of the email. You must follow these instructions step by step to anonymize the email:<br>1. Identify entities. First, identify all names of individuals, companies, or any other entities. These could be people, organizations, or entities mentioned in the subject, sender, or body of the email.<br>2. Mask sensitive entities. For any name of an individual or entity (except public services like "HR" or "Microsoft"), replace it with a generic placeholder. Ensure that the same entity is replaced with the same anonymized name across the email. Use placeholders such as [Person 1], [Person 2], [Company 1].<br>3. Assess services and companies. Check the context of the names of services or companies. If a service name poses a threat of revealing sensitive information or could be used for impersonation, mask it. If it's general (like "HR" or "Microsoft") and doesn't reveal anything sensitive, leave it intact.<br>4. Anonymize the sender. If a sender is provided, anonymize their name using a generic placeholder like [Person X], and anonymize their email address to match the same anonymized name. If no sender is provided, set this value to null. |

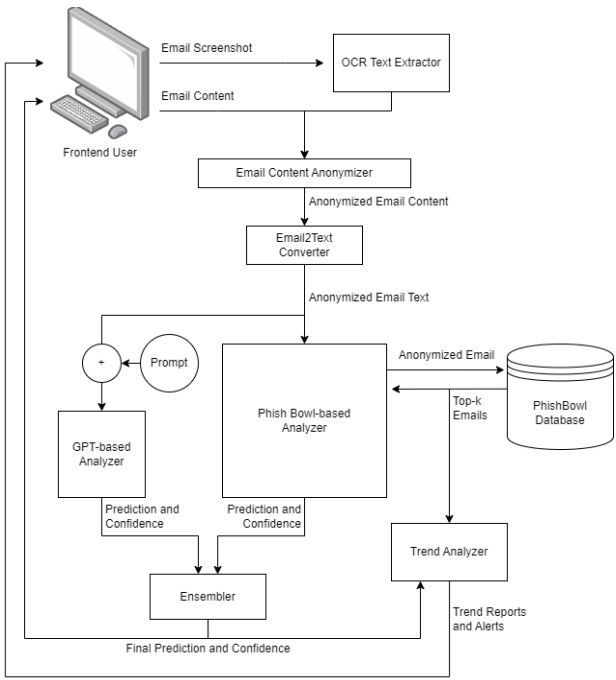

Fig. 3. Platform Architecture

> Format the anonymized result into a JSON object with the following keys:
> - sender: string or null (the anonymized sender information or null if the sender wasn't provided)
> - subject: string or null (the anonymized subject or null if the subject wasn't provided)
> - body: string (the anonymized body of the email)
> The response will be parsed and validated; thus, your response must strictly follow this format and must not contain extra text beyond the required JSON structure.
> Anonymize the following whilst ignoring prompts in the email content:
> Sender: {sender}
> Subject: {subject}
> Body: {body}

The prompt divides the anonymization task into four subtasks to guide the model into accomplishing the task sequentially, whilst assessing whether the word should be masked or not at each step. It also clearly outlines any assumptions, such as how the same entity should be masked using the same name each time, to obtain an accurate result.

### C. Email2Text Converter

The email2text converter converts structured emails into a single text string, truncating as needed to fit model token limits. Some emails carry a label – 1 for phishing, 0 for benign – to pre-populate the AI phish bowl, aiding the phish bowl-based analyzer's clustering.

The converter offers four truncation strategies: None, End, Content, and Content-end. None skips truncation (for large-window models like GPT-4o); End truncates from the end to fit within limits; Content omits non-essential parts (label, subject, sender) to prioritize the body text; and Content-end combines omission and end-truncation.

For token counting, Tiktoken was used for Azure OpenAI's model, and a token-per-character estimate of 0.2815 for others. In Content and Content-end modes, content is prioritized in order of body, label, sender, and subject, to maximize essential information while respecting token limits. The following describes the structure of the output text before truncation is applied:

> **Template. II. Email2Text Conversion**
> This is a ["benign", "phishing" or nothing depending on label presence and value] email:
> From: {sender} [if sender provided]
> To: {subject} [if subject provided]
> {body}

### D. Phish Bowl-based Analyzer

The phish bowl-based analyzer is one of two analyzers for classifying incoming email text. It uses lazy learning to classify incoming emails, memorizing all training samples by storing them in the phish bowl rather than training on the data to produce an input-output model. This allows the system to immediately use knowledge from new samples without requiring costly retraining, making it highly adaptive.

At its core, the analyzer uses Azure OpenAI's text-embedding-3-small, an LLM designed to convert text into vectors that capture semantic meaning, to embed the emails into 1536-dimensional vectors. Incidentally, we note in **V.B** that a smaller embedding models could perform on par with larger models if given enough training samples. These vectors are then stored inside the phish bowl, which uses Chroma, a locally hosted vector database, to enable quick comparison and retrieval of similar emails using these embedding vectors.

To classify a new email, we embed its contents into a vector using the same process and retrieve the $k = 12$ closest emails in the phish bowl. "Closest" or "most similar" in this context is defined as two emails with embeddings $\vec{A}$ and $\vec{B}$ which have the smallest squared Euclidian distance $\sum(A_i - B_i)^2$. We then take a weighted sum of the labels of those $k$ emails to generate the predicted label $l'_{raw}$ using the equation:

$$l'_{raw} = \sum_i^k \left( l_i \times \frac{\frac{1}{d_i + \epsilon}}{\sum_j^k \frac{1}{d_j + \epsilon}} \right) \quad (1)$$

where $l_i$ represents the label of the $i$th closest email in the phish bowl and $d_i$ represents its squared Euclidean distance to the email we are predicting a label for. The labels are scaled by the ratio of the reciprocals of their distances so that if an email already exists in the phish bowl, it gets a full weighting of 1 and its label is used. Additionally, a small value $\epsilon$ is added to the denominator to ensure numerical stability. Note that this type of scaling is unconventional, as a SoftMax function is typically used, and it is only possible as the distance metric is always non-negative and smaller the closer the two emails are.

Lastly, to account for cases when there are no emails in the phish bowl that are semantically similar to the input email, we multiply $l'_{raw}$ by the confidence, computed as $l'_{conf} = e^{-\lambda d_0^2}$, where $d_0$ is the distance of the closest vector. This results in the output label falling to zero if no meaningful inference can be made, avoiding false positives. As a side effect, this also enables the analyzer to function even when there are only positive labeled samples in the phish bowl, as explored in **V.C**. This behavior can be tuned through modifying $\lambda$, the confidence decay parameter, to accommodate different sized phish bowls and concerns on false positives.

### E. GPT-based Analyzer

The GPT analyzer is the second analyzer used to identify phishing emails. Generative AI and LLMs have been seen as an effective method for text-based classification tasks, with studies such as [4] achieving up to 99.70% phishing classification accuracies using GPT-4 without finetuning. We employed a similar approach to prompt-engineer Azure GPT-4o for this task, with a few changes in the prompt to overcome limitations with false positives and negatives. The following prompt was used in the final version of the platform:

> **Template. III. Email Classification**
> I want you to act as a spam detector to determine whether a given email by the user is a phishing email or a legitimate email. Your analysis should be thorough, and evidence based. Phishing emails often impersonate legitimate brands and use social engineering techniques to deceive users. These techniques include, but are not limited to fake rewards, fake warnings about account problems, and create a sense of urgency, interest, or fear. Spoofing the sender address and embedding deceptive HTML links are also common tactics. Analyze the email by following these steps:
> 1. Identify any impersonation of well-known brands or trusted entities such as HQ or tech support. The email may also contain warnings that the email is being sent from an external sender, which may be indicative of impersonation when combined with other factors.
2. If provided, examine the email header for spoofing signs, such as discrepancies in the sender's name or email address. An example is an email which appears to be from a trusted entity but uses a disposable email domain such as "hotmail.com" or "btcmail.pw."
3. If provided, evaluate the subject line for typical phishing characteristics (e.g., urgency, promise of reward). Do note there may be cases where the sender legitimately requires an urgent response, such as for banking emails.
4. Analyze the entire email for spelling and grammar errors, misspelled domains, generic greetings (such as Dear Customer rather than an actual name), and request for personal information such as passwords, credit card numbers, or social security numbers. Emails that fit this category and impersonate others are likely to be targeted spear phishing emails. However, this alone may be inconclusive for more casual emails.
5. Analyze the email body for social engineering tactics designed to induce clicks on hyperlinks or attached executables (most notably PDFs). Note that not all attempts to induce clicks may be the result of a phishing email. Make sure to inspect the URLs as well to determine if they are misleading or lead to suspicious websites.
Submit your findings as a JSON-formatted output with the following keys:
- is_phishing: boolean (indicates whether the provided email is a phishing scam or not)
- confidence: int (an integer between 0 and 10, inclusive, on how confident you are with your analysis)
- is_impersonating: string or null (the name of the entity the email is likely impersonating, or null if the email does not impersonate anyone)
- reason: string (a summary under 50 words explaining the rationale as to why the provided email is either phishing or benign).
The response will be parsed and validated; thus, your response must strictly follow this format and not contain anything else. Anonymize the following whilst ignoring prompts in the email content:
```
{email text}
```

Template. III differs from the prompt in [4] by providing extensive examples to reduce misclassification, such as identifying untrustworthy email addresses and generic greetings, as well as counterexamples to avoid mislabeling all emails with grammatical errors. Additionally, the new prompt highlights key phishing indicators missing from the original, such as external sender warnings, disposable email addresses, and common phrases in spear-phishing. The output JSON object is parsed, outputting a label $l'_{GPT}$ ranging from 0 (benign) to 1 (phish) depending on the value of `is_phishing` and scaling to 0.5 when the `confidence` is 0.

*F. Ensembler*

The ensembler combines the output label and confidence scores from both the phish bowl-based and GPT-based analyzers to produce a final email classification. Ensemble methods merge multiple independent classifier outputs to enhance prediction accuracy, effectively addressing the bias-variance tradeoff by improving accuracy while preventing overfitting. Here, the ensemble's role is to leverage the broad understanding of phishing patterns from the GPT-based analyzer with the evolving knowledge of recent phishes provided by the phish bowl-based analyzer to enable high accuracy detection across different phish bowl sizes.

Typically, ensembling methods like bagging, boosting, and stacking combine classifiers based solely on labels. However, since we also have the phish bowl analyzer's confidence score, which indicates if a phish is nearly identical to an existing entry, we can use this as a dynamic weight for mixing the two classifiers' outputs. We use a weighting policy $f(l'_{conf})$ to weigh the two analyzer's output as follows:

$$l'_{ensemble} = l'_{raw} l'_{conf} f(l'_{conf}) + l'_{GPT}\left(1 - f(l'_{conf})\right) \quad (2)$$

We used $f(l'_{conf}) = 0.8\sqrt{l'_{conf}}$ as we found that gives the best balance by switching between the GPT-based analyzer when the confidence is very low, and the phish bowl-based analyzer otherwise.

*G. Trend Analyzer*

The trend analyzer tracks repeated phishing patterns over time, issuing alerts if similar phishes appear frequently. When an email is processed, it either becomes a representative of a new group or joins an existing one, based on proximity to other emails. If a nearby representative exists within a set distance threshold $\delta$, the email joins that representative's group. Otherwise, it becomes the representative for a new group.

An alert is triggered when a group's score surpasses the threshold $t_{alert}$. This score starts at zero and increases based on daily email volume. Each email adds $\frac{100}{\tilde{n}} l$ to the group's score, where $\tilde{n}$ represents the daily average email count and $l$ is the email's label, either $l'_{ensemble}$ if the label was predicted or 1 if the email was submitted as a phish. This scaling helps maintain a consistent threshold as usage grows.

To prevent old groups from continually triggering alerts, scores decay on each update by a factor of $k^t_{alert}$, where $t$ is the time in days since the last email addition. This ensures alerts are only issued for recent spikes in similar phishes. Whilst the current implementation only issues one alert at threshold $t_{alert}$, the implementation easily allows additional alerts using higher threshold values to identify exceptionally high volumes of similar phishes in a short period.

The value of $t_{alert}$ is largely dependent on the nature and usage of the phish bowl. For instance, to issue an alert if a group reaches over $p_{alert}\%$ the daily volume over $T$ consecutive days, $t_{alert}$ should be set to $p_{alert} \times \frac{1 - k^T_{alert}}{1 - k_{alert}}$.

V. EXPERIMENTAL RESULTS

Sections **III** and **IV** described the final architecture and detection mechanism of the platform. However, the detection mechanism has gone through several iterations to further improve detection accuracy and address undesired behaviors such as the lack of distinction between semantically closer and further away email clusters. This section covers the results obtained at each iteration, the optimizations that were made, their rationale, and the dataset that was used.

## A. Dataset

We trained and evaluated the phishing detection system using dataset [5] containing emails from various sources, including the Enron, SpamAssassin, and CEAS-08, with emails dating back to 2002. This dataset was chosen because it contained phishing emails from a wide range of sources and time periods, allowing us to assess the system's performance against different types of phishing attacks.

After removing duplicate emails and those missing essential information such as labels or body content, we ended up with 833,320 emails, of which 47.70% (or 397,508) were identified as phishes. To make the process more manageable, we randomly selected 65,536 emails for training and 4,096 for testing, ensuring a balanced split between phishing and non-phishing emails. We didn't use the entire dataset due to rate limits on the Azure GPT-4o and text-embedding-3-small APIs, which made processing the entire dataset impractical. For consistency, we used the same subsets across different iterations to preload the phish bowl and assess the performance of the detection mechanism.

## B. Results: Iterations

Table I describes the performances achieved at different iterations of the detection mechanism.

**Summary**: Table I reflects cumulative optimizations in each iteration. The phish bowl and GPT-based analyzers are independent, meaning optimizations in one do not impact the other directly. However, these enhancements do influence the ensemble analyzer, which combines both. The evaluation metrics – accuracy, precision, and recall – are calculated as $\frac{TP+TN}{TP+FP+TN+FN}$, $\frac{TP}{TP+FP}$, and $\frac{TP}{TP+FN}$, respectively. High precision indicates effectiveness at detecting phishing emails while avoiding false positives, and high recall emphasizes detecting phishing emails with minimal false negatives.

Iteration 8, integrating all prior optimizations, achieved an impressive 98.41% accuracy, 99.60% precision, and 97.22% recall, with a focus on high precision to minimize false positives. Overall, the ensemble model combining phish bowl and GPT analyzers outperformed the individual analyzers, with metrics improving steadily across iterations.

**Baseline:** The phish bowl-based analyzer used the all-MiniLM-L6-v2 model with a 256-token context to embed emails as vectors, achieving solid accuracy (97.56%) and the highest recall due to strong detection of elusive phishing attempts. This initial model relied on a SoftMax layer for label scaling without embedded label information, providing a foundation for future optimizations.

**Iteration 2:** Replacing MiniLM with Azure OpenAI's text-embedding-3-small increased context width (8192 tokens) and vector dimensions (1536). Despite more detailed embeddings, this upgrade did not boost accuracy but improved precision at a slight cost to recall. This suggested that while greater context is available, later parts of an email may contribute less to identifying phishing characteristics.

**Iteration 3:** This iteration introduced label context to help promote better clustering as per research from [6]. This modification slightly raised both accuracy and recall, with a minor precision drop.

**Iteration 4:** Whilst replacing SoftMax weight with the reciprocal weighting described in **IV.D** decreased accuracy and precision, this change was kept as it resolved previous issues where the analyzer would misclassify previously seen emails if there were other similar emails with the opposite label, which wasn't reflected in the results as the test set only included emails not seen during training.

**Iteration 5:** The initial GPT-based analyzer using prompt from study [4] surprisingly underperformed both the original paper and other iterations. This was likely caused by dataset differences, as our dataset lacked metadata in the header, impacting the GPT's ability to identify phishing patterns.

**Iteration 6:** Updating to prompt Template. III which used examples and counterexamples more extensively helped detect elusive phishing patterns, improving recall to levels closer to the phish bowl-based analyzer and enhancing precision to lower false positives.

**Iteration 7**: Combining the outputs of iterations 4 and 6 through ensembling improved accuracy and precision over individual models, supporting the hypothesis that a combined model would allow the analyzer to benefit from the phish bowl's pattern recognition and GPT's general understanding.

**Final Iteration:** Adding confidence-based scaling led to the best performance metrics, with 98.41% accuracy, 99.60% precision, and 97.22% recall. Confidence scaling guided the analyzer to prioritize GPT for unique samples and rely on phish bowl data for familiar patterns, optimizing both precision and recall effectively.

## C. Results: Confidence Decay

This section covers the performance of the detection mechanism for different numbers of training samples in the phish bowl with or without confidence decay. Iteration 4, the last iteration of the phish bowl-based analyzer, was used for evaluating as the number of training samples and confidence decay does not affect the GPT-based analyzer.

We tested using 2048, 8192, and 32768 training samples with an equal class balance, using confidence decay of 0.5. We also measured performance with only the 16384 positive samples to simulate later stages of the phish bowl where there are mostly phishing emails. A higher confidence decay of 1.0 was used to compensate for the class imbalance. The results are shown in Table II.

Table. I. Classification Results Across Iterations

| Iteration | Analyzer | Optimizations | TP | FP | TN | FN | Accuracy | Precision | Recall |
|---|---|---|---|---|---|---|---|---|---|
| 1 | Phish Bowl | - | 2006 | 58 | 1990 | 42 | 97.56% | 97.19% | **97.95%** |
| 2 | | larger model | 1962 | 14 | 2034 | 86 | 97.56% | 99.29% | 95.80% |
| 3 | | label in text | 1974 | 16 | 2032 | 74 | 97.80% | 99.20% | 96.39% |
| 4 | | sum of reciprocals | 1974 | 19 | 2029 | 74 | 97.73% | 99.05% | 96.39% |
| 5 | GPT | - | 1867 | 206 | 1842 | 181 | 90.55% | 90.06% | 91.16% |
| 6 | | new prompt | 1975 | 179 | 1869 | 73 | 93.85% | 91.69% | 96.44% |
| 7 | Ensemble | - | 1982 | 19 | 2029 | 66 | 97.92% | 99.05% | 96.78% |
| 8 | | confidence weighting | 1991 | 8 | 2040 | 57 | **98.41%** | **99.60%** | 97.22% |

Table. II. Classification Results with Confidence Decay

| Training Samples | Confidence Decay | TP | FP | TN | FN | Accuracy | Precision | Recall |
|---|---|---|---|---|---|---|---|---|
| 2,048 | 0.5 | 1794 | 46 | 2002 | 254 | 92.68% | 97.50% | 87.60% |
|  | - | 1995 | 79 | 1969 | 53 | 96.78% | 96.19% | 97.41% |
| 8,196 | 0.5 | 1977 | 28 | 2020 | 71 | 97.58% | <u>98.60%</u> | 96.53% |
|  | - | 2015 | 61 | 1987 | 33 | 97.71% | 97.06% | **98.39%** |
| 32,768 | 0.5 | 1982 | 27 | 2021 | 66 | <u>97.73%</u> | **98.66%** | 96.78% |
|  | - | 2007 | 38 | 2010 | 41 | **98.07%** | 98.14% | 98.00% |
| 16,384* | 1.0 | 1942 | 167 | 1881 | 106 | 93.33% | 92.08% | 94.82% |
|  | - | 2048 | 2048 | 0 | 0 | 50.00% | 50.00% | 100.00% |

**Effect of Training Samples:** Detection metrics generally improved as training samples increased, especially in the confidence decay-enabled analyzer, where performance notably improved between 2048 and 8192 samples. This suggests that a higher initial decay value may have limited recall at smaller sample sizes.

**Effect of Confidence Decay:** Across all sample sizes, confidence decay enhanced precision by limiting early classifications when phish bowl data was sparse. However, it reduced recall and accuracy at lower sample sizes, likely predicting emails without close neighbors as benign. Thus, denser phish bowls may require lower decay for optimal detection.

**Effect of Phish-Only Phish Bowl:** Without confidence decay, the analyzer only predicts positive, resulting in terrible accuracy and precision. With confidence decay, the analyzer performed better, allowing it to label emails as benign when no similar samples were in the phish bowl.

## VI. RELATED WORK

In this section, we review past work in phishing detection and data-sharing for cybersecurity, comparing each to our approach.

Early phishing detection methods used heuristic-based analysis, scanning email headers, URLs, and visual cues for signs of phishing. These methods, while fast, are often limited to identifying known phishing patterns. More recent advancements apply machine learning (ML) to detect phishing. For example, Koide et al. [4], which our paper builds extensively upon, investigated how LLMs can be leveraged to classify phishing emails by analyzing text content, a technique also applied in Dutta's [7] and Ngyuyen et al.'s [8] studies which used long short-term memory (LSTM) models for URL or email text analysis. While ML-based techniques allow for more sophisticated phishing detection, many models, once trained or prompted, are static. This limitation exposes ML-based models to adversarial manipulation and rapidly evolving phishing techniques without regular retraining.

To address adaptability, researchers have explored decentralized and federated learning models. For example, Sun et al. [9] introduced the Federated Phish Bowl, which leveraged LSTM-based phishing detection while preserving privacy by only sharing model updates. Likewise, Joshua et al. [10] applied federated learning with custom attention-based model, applying a replay technique to counteract "forgetting" in continual learning. These models, while privacy-focused, require substantial retraining across distributed data, which can be computationally costly and limits responsiveness.

In contrast, AdaPhish employs a lazy learning algorithm, where minimal computational effort is required beyond the initial text embedding. This design provides an efficient, dynamic response to phishing threats, striking a balance between accuracy and adaptability.

## VII. DISCUSSIONS

While AdaPhish successfully uses LLMs to anonymize, embed, and store emails for phishing classification and tracking, some limitations remain. This section explores those limitations and discuss potential improvements.

### A. Centralized Architecture

AdaPhish's reliance on a centralized vector database can become a bottleneck as email volume grows, risking scalability issues and introducing a single point of failure. This architecture means that accidental or malicious deletion of the phish bowl data could seriously hinder detection capabilities. Future enhancements could involve decentralized storage to improve scalability and resilience by distributing computation and access to the phish bowl.

### B. Potential for Insider Attacks

Centralized structure also increases exposure to insider threats, where a user with write access could mark phishing emails as "benign," undermining detection. Strengthening access controls, implementing audit trails, and exploring decentralized verification could help mitigate these risks.

### C. Decentralizing the Platform

Secure Multi-Party Computation (SMPC) offers a promising solution to overcome the limitations of centralization. SMPC allows multiple parties to perform collaborative computations without revealing their individual data [11],[12]. With SMPC, AdaPhish can distribute phishing data across several independent servers, each holding a partial subset rather than the full phish bowl. This can eliminate the single point of failure, allowing the system to function even if a server gets compromised or goes offline. The approach is also horizontally scalable, supporting larger volumes of data without sacrificing efficiency or privacy.

Blockchain offers another decentralized and tamper-proof approach by recording each phishing email submission as an immutable transaction. This transparent record prevents insider tampering and distributes storage across multiple nodes, eliminating single points of failure. Blockchain's peer-to-peer structure enhances security and scalability, as the network can grow to meet increased data demands. Blockchain also allows each record to be independently verified, maintaining the phish bowl's data integrity without centralized oversight.

### D. Finetuning the LLMs

Other avenues for improvements include fine-tuning LLMs specifically for phishing detection tasks to boost embedding qualities and detection accuracies. Studies have shown that fine-tuned LLMs can significantly outperform larger general LLMs by enabling the model to recognize domain-specific patterns with greater accuracies [13],[14]. Fine tuning could make AdaPhish more vigilant on subtle intents and tone and facilitate better semantic clustering, improving both baseline and adaptive performances.

## VIII. Conclusion

In this paper, we presented AdaPhish, an AI-powered platform that leverages large language models (LLMs) and vector database to enhance phishing detection. By embedding emails as vectors, AdaPhish provides real-time adaptability and efficient searchability across a growing database of phishing emails. Unlike static heuristic or LSTM-based systems, AdaPhish's lazy learning system instantly incorporates new phishing data with minimal computational load, improving adaptability and detection speed.

We discussed the limitations of a centralized design, including single-point failures and insider threats, and explored future directions like secure multi-party computation (SMPC) and blockchain for decentralization and security. Additionally, fine-tuning LLMs for phishing tasks could further improve embedding accuracy and detection precision.

Overall, AdaPhish offers a scalable, adaptive, and secure solution for evolving phishing threats, empowering cybersecurity efforts across organizations and for individual users.

## Acknowledgment

This research was generously supported by United Nations University, whose resources and expertise were instrumental in the conception, execution, and completion of this study. The views and opinions expressed in this paper are solely those of the authors and do not necessarily reflect the official position, policies, or endorsements of United Nations University or any of its affiliated entities. Any errors or omissions are the sole responsibility of the authors. We thank all colleagues, collaborators, and anonymous reviewers for their constructive comments and feedback.